\documentclass[aps,prl,twocolumn,showpacs,floatfix,preprintnumbers,nofootinbib,superscriptaddress]{revtex4}

\usepackage{graphicx}
\usepackage{color}
\usepackage{natbib}
\usepackage{multirow}
\usepackage{amsmath}
\usepackage[amssymb]{SIunits}

\newcommand{\vek}[1]{\boldsymbol{#1}}
\newcommand{\dNeff}{\Delta N_{\textrm{eff}}}
\newcommand{\fref}[1]{Fig.~\ref{#1}}

\begin{document}

\title{How secret interactions can reconcile sterile neutrinos with cosmology}

\author{Steen Hannestad}
\affiliation{Department of Physics and Astronomy,
 University of Aarhus, 8000 Aarhus C, Denmark}

\author{Rasmus Sloth Hansen}
\affiliation{Department of Physics and Astronomy,
 University of Aarhus, 8000 Aarhus C, Denmark}

\author{Thomas Tram}
\affiliation{Institut de
Th\'eorie des Ph\'enom\`enes Physiques, \'Ecole Polytechnique
F\'ed\'erale de Lausanne, CH-1015, Lausanne,
Switzerland}

\date{\today}

%\preprint{blahblah}

\begin{abstract}
  Short baseline neutrino oscillation experiments have shown hints of the existence of 
	additional sterile neutrinos in the eV mass range. However, such neutrinos
	seem incompatible with cosmology because they have too large an impact 
	on cosmic structure formation. Here we show that new interactions in the sterile
	neutrino sector can prevent their production in the early Universe and
	reconcile short baseline oscillation experiments with cosmology.
\end{abstract}


\maketitle

{\em Introduction.}---A variety of short baseline neutrino experiments 
seem to indicate the existence of at least one more neutrino species
with a mass in the eV range (see e.g.\ \cite{Kopp:2013vaa,Giunti:2012bc}). In order to be compatible with the LEP
constraint on the number of light neutrinos coupled to $Z$ \cite{ALEPH:2005ab} these
additional neutrinos must be sterile, i.e.\ they must be singlets under the
$SU(2) \times U(1)$ electroweak gauge group. However, the fact that they do not
couple to any particles in the standard model by no means 
implies that they are completely non-interacting.
In fact it is entirely possible, even natural that the sterile neutrinos couple to
other vector bosons which can have different properties from those associated with the
$SU(2)\times U(1)$ of the standard model. Here we will consider the possibility that sterile neutrinos 
can be strongly self-coupled through a ``secret'' Fermi 4-point interaction similar to the low energy
behaviour of neutrinos in the standard model, but with a completely different 
coupling strength. As we will see below such a new interaction can have profound effects on active-sterile neutrino
conversion in the early Universe and completely change cosmological bounds on sterile neutrinos.

Recent data on the anisotropy of the cosmic microwave background (CMB) from the Planck satellite \cite{Ade:2013zuv}, in combination with auxiliary data on the 
large scale distribution of galaxies has shown that
cosmology can accommodate sterile neutrinos of eV mass, but {\it not} if they are fully thermalized (see e.g.\ \cite{Archidiacono:2013xxa}) because the suppression of
structure formation is too strong (see \cite{Hannestad:2010kz,Wong:2011ip,Lesgourgues:2012uu} for a detailed discussion).
The problem is that the masses and mixing angles preferred by terrestrial data inevitably leads to almost 
complete thermalization of sterile neutrinos. One possible way of circumventing this problem is to
introduce a lepton asymmetry which pushes the resonant region in momentum space to very low values (see e.g.\ \cite{Hannestad:2012ky,Saviano:2013ktj}).
The problem with this model is that it is far from clear how to produce this lepton asymmetry.
Furthermore, the suppression changes very rapidly from zero to maximum suppression as a function of the lepton asymmetry, so partial
thermalization requires some fine-tuning.
Here we present an alternative scenario for preventing sterile neutrino production: If sterile neutrinos
are strongly self-interacting  they provide a significant matter potential for themselves which in turn completely changes the active-sterile conversion process.
We will demonstrate that self-interactions can prevent sterile neutrino production to a point where bounds from CMB and large scale structure completely
disappear - making sterile neutrinos with masses in the eV range perfectly compatible with precision cosmological data.

{\it Scenarios.}---
We are considering a hidden gauge boson with mass $M_X$, and we take the mass to be $\gtrsim 100\text{MeV}$ such that we can use an effective 4-point interaction for all temperatures of interest. The interaction strength is then written as
\begin{equation}
G_X \equiv \frac{g_X^2}{M_X^2}. %Rasmus, do you take the usual factors in this definition?
% Answer: I include them in g_X, so this definition is what we have used.
\end{equation}
We will assume a 1+1 scenario, specifically a muon neutrino (or tau neutrino) and 1 sterile neutrino species, a simplification which does not qualitatively alter any of our findings. The system can then be fully characterized by a momentum dependent, $2\times 2$ Hermitian density matrix $\rho(p)$. Since we are not assuming any lepton asymmetry, the evolution of the anti-particle density matrix is trivial, since $\rho(p) = \bar{\rho}(p)$. We expand the density matrix in terms of Pauli matrices:
\begin{equation}
\rho = \frac{1}{2} f_0 (P_0 + \vek{P} \cdot \vek{\sigma}),
\end{equation}
where $f_0=(e^{p/T}+1)^{-1}$ is the Fermi-Dirac distribution and $\vek{\sigma}$ is a vector consisting of the three Pauli matrices. The evolution equations for $P_0$ and $\vek{P}$ are called the quantum kinetic equations (QKE), and they were first derived in~\cite{Barbieri:1990vx,Enqvist:1990ad,Sigl:1992fn,McKellar:1992ja} (for a presentation closer to the present one, see~\cite{Kainulainen:2001cb,Hannestad:2012ky}). It is convenient to form the linear combinations:
\begin{subequations}
\begin{align}
P_a \equiv P_0 + P_z = 2\frac{\rho_{aa}}{f_0}, \\
P_s \equiv P_0 - P_z = 2\frac{\rho_{ss}}{f_0},
\end{align}
\end{subequations}
which separates the sterile and the active sector. The equations of motions are then given by
\begin{subequations}
\begin{align}
\dot{P_a} &= V_x P_y + \Gamma_a \left[2\frac{f_0}{f_0}-P_a \right], \\
\label{eq:Psdot}
\dot{P_s} &= -V_x P_y +\Gamma_s \left[2 \frac{f_{\text{eq},s}(T_{\nu_s},\mu_{\nu_s})}{f_0} - P_s \right], \\
\dot{P_x} &= -V_z P_y - D P_x, \\
\dot{P_y} &= V_z P_x -\frac{1}{2}V_x(P_a-P_s)- D P_y.
\end{align}
\end{subequations}
The $\Gamma_s$-term is an approximation to the full scattering kernel which is valid in the limit of strong coupling. The sterile equilibrium distribution:
\begin{equation}
f_{\text{eq},s}(T_{\nu_s},\mu_{\nu_s}) = \frac{1}{e^{(p-\mu_{\nu_s})/T_{\nu_s}}+1},
\end{equation}
where $T_{\nu_s}$ and $\mu_{\nu_s}$ are the sterile neutrino temperature and pseudo-chemical potential respectively,
is uniquely determined from the requirement that the interaction must respect energy conservation and number conservation. $\Gamma_a$ and $\Gamma_s$ are related to the 4-point interaction constants as
\begin{equation}
\Gamma_a = C_\mu G_F^2 p T^4,\qquad \Gamma_s = G_X^2 p T_{\nu_s}^4 n_{\nu_s},
\end{equation}
where $C_\mu \simeq 0.92$, while $n_{\nu_s}$ is the normalized number density of sterile neutrinos, $n_{\nu_s} = \frac{2}{3\zeta(3)T^3}\int p^2 \rho_{ss}(p) dp $. $D$ quantifies the damping of quantum coherence in the system and is approximately half of the scattering rates, $D\simeq \frac{1}{2}(\Gamma_a+\Gamma_s)$. 
We have chosen to define $\Gamma_s$ in analogy with $\Gamma_a$, and this means that we do not have exact conservation of $\dNeff$ for the scattering term in Eq.~(\ref{eq:Psdot}) since $\Gamma_s$ depends on $p$. However, none of the results change significantly when we let $p=3.15T$ in the expression for $\Gamma_s$.
%With the above expression for $\Gamma_s$ we do not have explicitly that $d\dNeff/dt=0$ for the scattering term in Eq.~(\ref{eq:Psdot}), but this does not affect any of our results significantly.

In order to include the sterile neutrino self-interaction, we repeat the derivation in~\cite{Sigl:1992fn} for the self-interaction due to the $Z$-boson in the active sector, but now for an $X$-boson in the sterile sector. This gives an addition to the matter-potential $V_z$. The potentials are now
\begin{subequations}
\begin{align}
V_x &= \frac{\delta m_s^2}{2p} \sin 2\theta, \\
%V_y &= 0, \\
V_z &= V_0 + V_a + V_s, \\
V_0 &= -\frac{\delta m_s^2}{2 p} \cos 2\theta, \\
V_a &= -\frac{14\pi^2}{45\sqrt{2}} p \left[\frac{G_F}{M_Z^2} T_\gamma^4 n_{\nu_a} \right], \\
V_s &= +\frac{16 G_X }{3\sqrt{2} M_X^2}p u_{\nu_s} .
%V_s &= +\frac{8 G_X}{3 \sqrt{2} \pi^2 M_X^2} p T_\gamma^4 \int q^3 f_0 P_s dq , \quad q = p/T .
%V_s &= +\frac{16 G_X}{3 \sqrt{2} \pi^2 M_X^2} p T_\gamma^4 \int q^3 \rho_{ss} dq , \quad q = p/T .
\end{align}  
\end{subequations}
Here $\delta m_s^2$ is the mass difference, $\theta$ is the vacuum mixing angle, $M_Z$ is the mass of the Z-boson, $M_X$ is the mass of the boson mediating the secret force, and $u_{\nu_s}$ is the physical energy density of the sterile neutrino. We solve the system of equations using a modified version of the public code \texttt{LASAGNA}~\cite{Hannestad:2013pha} available at~\url{http://users-phys.au.dk/steen/codes.html}.

\begin{figure}[tb]
\includegraphics[width=7.5cm]{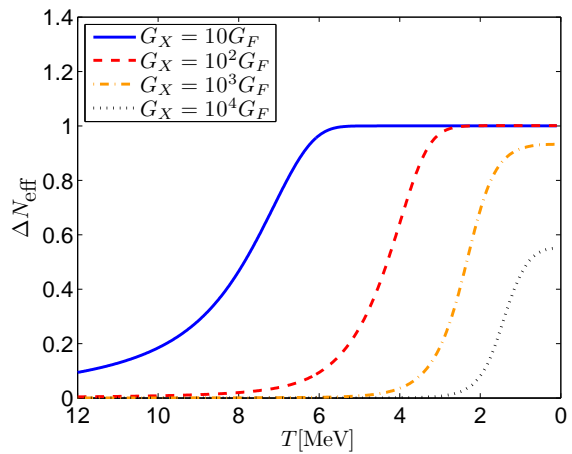}
\caption{The evolution of $\dNeff$ as the temperature drops for $g_X = 0.1$ and different values of the coupling constant $G_X$.\label{fig:temp}}
\end{figure}

{\it Results.}---
In \fref{fig:temp} we show the degree of thermalization of the sterile neutrino, quantified in terms of the total energy density in the active plus sterile sector,
\begin{equation}
N_{\rm eff} \equiv \frac{u_{\nu_a}+u_{\nu_s}}{u_{\nu_0}} \,\, , \,\, u_{\nu_0} \equiv \frac{7}{8}\left(\frac{4}{11}\right)^{4/3} u_\gamma.
\end{equation}
We have chosen $g_X=0.1$ and a sample of values for $G_X$, and we show how $\dNeff$ develops with the decreasing temperature. We can see that the thermalization of the sterile neutrino moves to lower temperatures when the interaction becomes stronger, and this is what we would expect since a strong interaction means that even a small background of sterile neutrinos can prevent further thermalization.

\begin{figure}[tb]
\includegraphics[width=7.5cm]{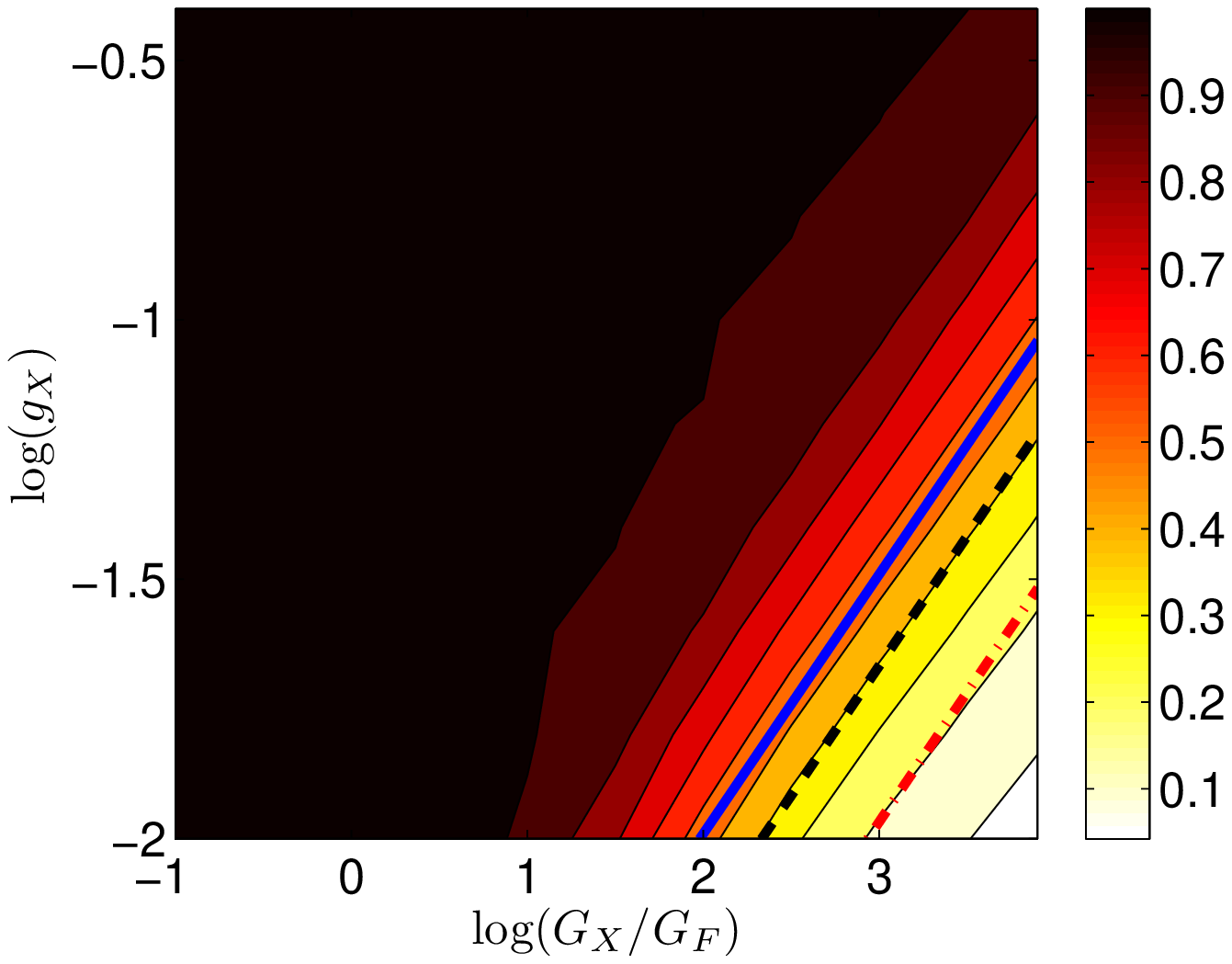}
\caption{Contours of equal thermalization. $\dNeff$ is given by the colors. The solid, dashed, and dot-dashed lines correspond to hidden bosons with masses $M_X = 300\:\mega\electronvolt$, $200\:\mega\electronvolt$, and $100\:\mega\electronvolt$ respectively.\label{fig:contours}}
\end{figure}

The amount of thermalization depends on both $g_X$ and $G_X$, and in \fref{fig:contours} we show $\dNeff$ as a function of both.
It shows that thermalization can be almost completely blocked by the presence of the new interaction for high values of $G_X$ and low values of $g_X$. 

Another interesting observation is that the degree of thermalization depends almost entirely on the mass of the new boson, $M_X$, not on the dimensionless coupling $g_X$. This can be understood qualitatively from the following simple argument: 
At high temperature the production of sterile neutrinos is suppressed by rapid scattering (the quantum Zeno effect), but as soon as production commences the thermalization rate of a sterile neutrino can be approximated by 
\begin{equation}
\Gamma_t \sim \Gamma \sin^2 (2\theta_m),
\end{equation}
where $\Gamma$ is the rate with which ``flavor content'' (in this context meaning active vs. sterile) is measured by the system and $\theta_m$ is the in-medium mixing angle (see e.g.\ \cite{Stodolsky:1986dx,Hannestad:1999zy} for a discussion of this in the context of active neutrinos). $\Gamma$ is entirely dominated by the interaction via $X$ so that $\Gamma \propto G_X^2$ and the in-medium mixing angle is likewise dominated by the potential generated by the new interaction so that $\sin^2 (2\theta_m) \propto 1/V_s^2 \propto M_X^4/G_X^2$ leading to the sterile thermalization rate being proportional to $M_X^4$, i.e.\ $\Gamma_t$ does not depend on $g_X$, only on $M_X$.

\begin{figure}[tb]
\includegraphics[width=7.5cm]{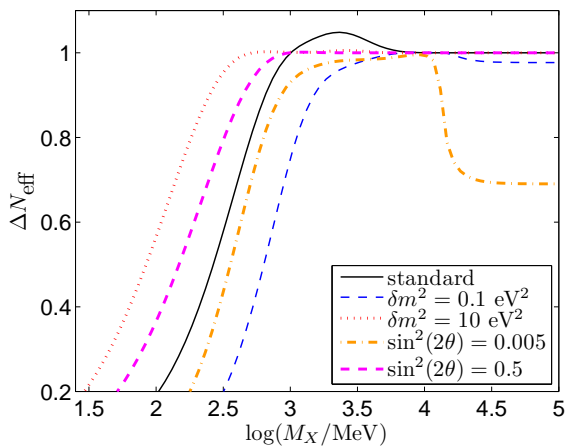}
\caption{Dependence of $\dNeff$ on the mixing parameters. $g_X = 0.01$ has been used for all the models while $G_X$ has been changed to give the variation in mass. 
%Note that the thermalization is not complete for $\sin^2(2\theta) = 0.005$, when the hidden boson mass is large as it is expected~\cite{Hannestad:2012ky}.
\label{fig:mass}}
\end{figure}

The determination of mixing parameters from accelerator experiments is quite uncertain, and it is therefore interesting to know how our results would be affected if we changed the vacuum mixing angle or the mass difference. The results of such a variation are seen in \fref{fig:mass}.
Regarding the ability to inhibit thermalization, the results do not change much. A somewhat higher or lower mass will be needed for the hidden boson, but $\dNeff = 0.6$ can for example be reached by using $M_X = 100\:\mega\electronvolt$ even at $\delta m^2 = 10\:\electronvolt^2$.
There are, however, two other interesting observations. First, note that when the hidden boson mass is high, $\dNeff$ decreases with decreasing $\sin^2(2\theta)$ or $\delta m^2$ - the well known limit for non-interacting sterile neutrinos (see e.g.\ \cite{Enqvist:1991qj,Hannestad:2012ky}). As the boson mass is lowered, the new interaction first permits full thermalization of the sterile neutrino before we reach the mass range where the new interaction inhibits the thermalization.

\begin{figure}[tb]
\includegraphics[width=7.5cm]{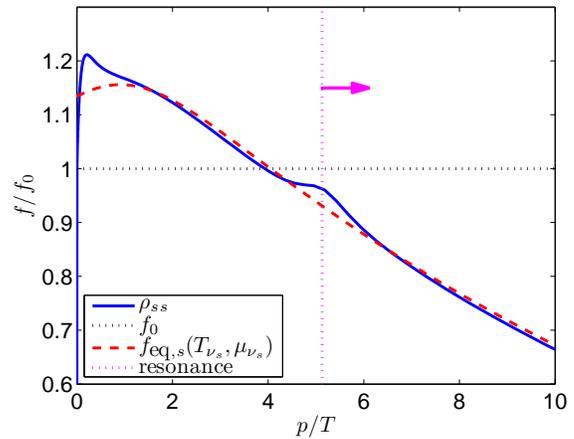}
\caption{
%Upper panel: The sterile energy distribution at $T=4.3\:\mega\electronvolt$, where $\dNeff = 1$. 
%Lower panel: 
The sterile energy distribution relative to $f_0$ at $T=4.3\:\mega\electronvolt$, where $\dNeff$ crosses 1 for $\delta m^2 = 1\:\electronvolt^2$, $\sin^2(2\theta) = 0.05$, $G_X = G_F$, and $g_X = 0.01$ which corresponds to $M_X = 2.9\:\giga\electronvolt$. Note that the peak at $p/T < 1$ is unimportant due to the limited phase space for so low $p$.
%{\bf Vil vi have noget om processen i caption? Nej, men parametrene skal st{\aa} der, dvs $G_X$, $\sin^2(2\theta)$ og $\delta m^2$}
%$\dNeff$ is still growing since the active neutrino distribution, which follows $f_0$ quite closely is larger at the resonant momentum, while the equilibrium distribution $f_eq(T_x,\mu_s)$ is larger the the sterile distribution between $p/T = 2$ and $4$. The range for small $p/T$ where the sterile distribution is larger than $f_eq(T_x,\mu_s)$ is suppressed in the number integral by a factor of $(p/T)^2$.
\label{fig:Neff_gt_1}}
\end{figure}

The other interesting observation is that $\dNeff>1$ for some values of $M_X$. At first this seems very puzzling and counterintuitive. In a model with only oscillations and no new interactions this would be impossible since the number density and energy density of the sterile neutrinos could never exceed the densities of the active neutrinos, the net production of steriles would simply shut off as soon as $\rho_{ss} \sim f_0$. However, in the model presented here there are two effects at play simultaneously: The production of steriles due to oscillations and the redistribution of sterile states due to the new interaction. If the redistribution of energy is sufficiently fast it can keep $\rho_{ss} < f_0$, allowing for more production of steriles. \fref{fig:Neff_gt_1} provides an illustration of the effect by showing a snapshot of the distributions at the point where $\dNeff$ crosses 1 for a model with $M_X = 2.3\:\giga\electronvolt$.
Sterile neutrinos are still being produced in the region close to the resonance at $p/T \approx 5$ since $f_0 > \rho_{ss}$ and oscillations therefore populate sterile neutrinos from the active sector. At the same time $\rho_{ss}$ continues to grow at lower $p/T$ due to the redistribution of states.
In total this means that $\dNeff$ is still growing and will do so until  the resonance has moved to very high $p/T$ where $f_0$ becomes very small or the active neutrinos decouple from the electrons. 
Naively we would expect $\dNeff$ to be highest for low values of $M_X$ because the energy redistribution becomes more efficient. However, when $M_X$ is decreased the suppression of oscillations due to the effect of $M_X$ on the matter potential quickly wins and $\dNeff$ decreases rapidly with decreasing $M_X$. Therefore $\dNeff > 1$ can only occur in a limited transition region of $M_X$ if it occurs at all (which depends on the mixing parameters, $\delta m^2$ and $\sin^2(2\theta)$).

Finally, we again stress that our treatment is only consistent if $M_X \gg T$ for {\it any} temperature relevant to our calculation. For the typical mass differences favoured by SBL measurements the production of sterile neutrinos takes place at temperatures well below $100\:\mega\electronvolt$ and we have taken this as a representative minimum mass for the new boson. Note that such a low mass would be completely excluded for a boson coupling to the active sector \cite{Bilenky:1999dn}. However, provided that the coupling is diagonal in ``flavor'' such that $X$ couples only to the sterile state, such bounds are irrelevant. 

\begin{figure}[tb]
\includegraphics[width=7.5cm]{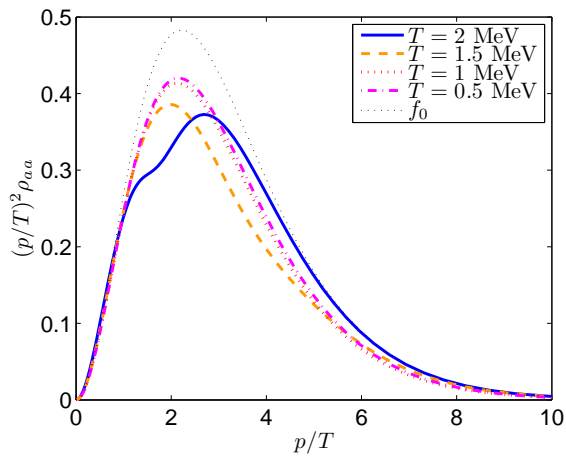}
\caption{
%Upper panel: The active neutrino distribution for different temperatures. The parameters used are $G_X = 3\cdot10^3G_F$ and $g_X=0.025$. This corresponds to a hidden boson with the mass $M_X = 134\mega\electronvolt$. 
%Lower panel: 
The active neutrino distribution for different temperatures. The parameters used are $G_X = 3\cdot10^2G_F$ and $g_X=0.025$. This corresponds to a hidden boson with the mass $M_X = 424\:\mega\electronvolt$. \label{fig:dist}}
\end{figure}

{\it Big Bang Nucleosynthesis (BBN).}---
Apart from the additional energy density in the sterile sector the oscillations can have another important effect, namely a distortion of the active neutrino distribution. This can happen even after neutrino decoupling because energy can still be transferred between the active and sterile sectors after the active neutrino decouples from the plasma.
In models where the active-sterile conversion is delayed, such as the one presented here or models with a non-zero lepton asymmetry \cite{Saviano:2013ktj} this can in certain cases be the dominant cosmological effect. The reason is that the electron neutrino takes part in the nuclear reaction network relevant for Big Bang Nucleosynthesis (see e.g.\ \cite{Saviano:2013ktj}). Even if the sterile neutrino mixes primarily with $\nu_\mu$ or $\nu_\tau$, active-active oscillations will transfer part of the distortion to the electron sector. However, a detailed investigation of this effect is beyond the scope of the present paper and here we simply point out that interesting effects on BBN might occur. For illustration we show in \fref{fig:dist} how the active distribution can vary as a function of temperature relative to its unperturbed state, $f_0$.

{\it Discussion.}---
We have demonstrated that additional self-interactions of a sterile neutrino can prevent its thermalization in the early Universe and in turn make sterile neutrinos compatible with precision cosmological observations of structure formation. Arguably the model discussed here is more natural than invoking a non-zero lepton asymmetry, relying only on the sterile sector possessing interactions similar to those in the standard model. In order for the model to work the new gauge boson mediating the interaction must be significantly lighter than $M_Z$, but can easily be heavy enough that no significant background of such particles can exist at late times. We finally note that if this scenario is indeed realized in nature, future precise measurements of $N_{\textrm{eff}}$ will effectively pinpoint the mass of the hidden gauge boson.
In summary, the framework presented here presents a natural way of reconciling short baseline neutrino experiments with precision cosmology.

{\it Acknowledgments.}---We thank Georg Raffelt for valuable comments on the manuscript.

%%%%%%%%%%%%%%%%%%%%%%%%%%%%%%%%%%%%%%%%%%%%%%%%%%%%%%%%%%%%%%%%%%%%%%%

%%%%%%%%%%%%%%%%%%%%%%%%%%%%%%%%%%%%%%%%%%%%%%%%%%%%%%%%%%%%%%%%%%%%%%%
\end{document}